\documentclass[aps,twocolumn,a4paper,pra,superscriptaddress,floatfix]{revtex4}%
\usepackage{psfig}
\usepackage{amssymb}
\usepackage{amsmath}
\usepackage{amsfonts}
\usepackage{graphicx}%
\def\P{\mathrm{P}}    
\def\C{\mathrm{C}}    
\def\PP{\mathrm{PP}}  
\def\PS{\mathrm{PS}}  
\def\lat{\mathrm{lat}}  
\def\br{\mathbf{r}}
\def\bk{\mathbf{k}}
\def\dd{\mathrm{d}}

\begin{document}
\title{Lateral Casimir force beyond the Proximity Force Approximation}
\author{Robson B. Rodrigues}
\author{Paulo A. Maia Neto}
\affiliation{Instituto de F\'{\i}sica, UFRJ, 
CP 68528,   Rio de Janeiro,  RJ, 21941-972, Brazil}
\author{Astrid Lambrecht}
\author{Serge Reynaud}
\affiliation{Laboratoire Kastler Brossel,
CNRS, ENS, Universit\'e Pierre et Marie Curie case 74,
Campus Jussieu, F-75252 Paris Cedex 05, France}

\date{\today}

\begin{abstract}
We argue that the appropriate variable to study a non trivial geometry dependence of the Casimir force is the lateral component of the Casimir force, which we evaluate between two corrugated metallic plates outside the validity of the Proximity Force Approximation (PFA). 
The metallic plates are described by the plasma model, with arbitrary values for the plasma wavelength, the plate separation and the corrugation period, 
the corrugation amplitude remaining the smallest length scale. Our analysis shows that in realistic experimental situations the Proximity Force Approximation overestimates the force by up to 30\%. 
\end{abstract}
\maketitle

Considerable experimental progress has been achieved~\cite{review} in the measurement 
of the Casimir force, opening the way for various applications in nano-science~\cite{capasso}, 
particularly in the development of nano- or micro-electromechanical systems (NEMS or MEMS). 
Calculations are much simpler in the original Casimir geometry of two plane plates~\cite{Casimir}
which obeys a symmetry with respect to lateral translations and thus allows
to derive the expression of the Casimir force from the reflection amplitudes which
describe specular scattering on the plates~\cite{plane_plane}.

More general geometries open a far richer physics with a variety of 
extremely interesting theoretical predictions~\cite{geometry}. 
Up to now the experimental studies of the effect of geometry have been
restricted to simple configurations which can be calculated 
with the help of the Proximity Force Approximation (PFA).
This approximation is essentially equivalent to an averaging over plane-plane geometries
and its result can be deduced from the force known in this geometry~\cite{PFA}. 
For example, it allows to evaluate the force between a plane and a sphere~\cite{plane_sphere}   
provided the radius $R$ of the sphere is much larger than the mirror separation $R\gg L$.
It is also valid for the description of the effect of roughness when the wavelengths
associated with the plate deformations are large enough~\cite{PFA2}. However PFA relies heavily on assuming some additivity of Casimir forces which is known to be generally not valid except for very smooth geometrical perturbations~ \cite{EPL}.

The aim of the present paper is to study a configuration allowing a new test 
of QED theoretical predictions outside the PFA domain and independent of those already performed in the plane-plane geometry.
The idea is to look for the lateral component of the Casimir force which appears, 
besides the usual normal component, when periodic corrugations with the same 
period are imprinted on the two metallic plates. This configuration contrasts with other ones, for example the normal Casimir force in the plane-sphere geometry or roughness corrections to it. There PFA can also be invalid, but this leads only to small corrections of the dominant normal Casimir force, which do not seem accessible experimentally at the moment.
The lateral component of the Casimir force has recently been measured and analyzed within the PFA~\cite{chen,blagov}. 
We find for experimentally realizable parameters that PFA overestimates the force by as much as 30\%, which should allow for a rapid experimental check of its validity. 

The lateral Casimir force between corrugated surfaces has been analyzed outside the PFA domain for perfect reflectors~\cite{emig} where
interesting results were obtained for arbitrary values of the ratio $\lambda_\C/L$ 
of the corrugation wavelength $\lambda_\C$ to the interplate distance $L$. 
However, the assumption of perfect reflections can only be valid in the limit of large distances where the lateral force tends to become too weak to be measurable.
Experimental conditions allowing the lateral force to be measured correspond to 
separations of a few hundred nanometers~\cite{chen}, that is of the same order of magnitude as the plasma wavelength $\lambda_\P$ associated with the metallic plates, so that they cannot be treated even as approximately perfect reflectors~\cite{metals}. In contrast, the calculations presented here allow to predict the lateral force in this domain.

In this letter, we calculate the lateral force for metallic plates modeled by the plasma model
with arbitrary values of $L,$ $\lambda_\C$ and $\lambda_\P.$
We use the perturbative approach that we developed for analyzing the effect of 
stochastic roughness on the normal Casimir force~\cite{MLR,MLR2}. 
This technique is valid as long as the corrugation amplitude $a$ remains
the smallest of length scales $a\ll L,\lambda_\C,\lambda_\P$.
As this condition does not depend on the relative magnitudes of the three other length scales,
it allows us to study situations beyond the PFA where the lateral force is experimentally
accessible. The result will be expressed in terms of a non linear susceptibility function,
calculated from non-specular scattering amplitudes \cite{non-specular} associated with corrugated metallic surfaces.
This function, obtained within the scattering approach \cite{lossy_cavities}, can itself be considered as a new QED theoretical prediction to be compared with forthcoming experiments. 

The surface profiles of the two parallel corrugated plates are defined by two functions 
$h_{1} (\br) $ and $h_{2} (\br),$
with $\br=(x,y)$  the lateral position along the surfaces of the plates.
Both distributions $h_{1}$ and $h_{2}$ have zero spatial averages $\langle h_{j}\rangle =0,j=1,2,$ 
and they are counted as positive when they correspond to local length decreases below the mean value $L$. 
The corrugated surfaces are assumed to be static, so that the field frequency $\omega$ is preserved by 
scattering. In contrast, the lateral wavevector components $\bk=(k_x,k_y)$ as well as the 
polarization of the field are modified. 
Scattering on the plate $j=1,2$ is thus described by non-specular reflection amplitudes
${\cal R}_{j;pp'}(\bk,\bk',\omega),$
where $\bk,\bk'$ represent the lateral wavevectors of the input and output fields 
and $p$ and $p'$ their polarizations, TE for transverse electric and TM for transverse magnetic
(more detailed definitions in~\cite{MLR,MLR2}). 

For the purpose of the present paper, the non-specular reflection amplitudes
have to be developed up to the first order in the deviations $h_j$ from flatness
of the two plates
\begin{eqnarray}
{\cal R}_{j;pp'}(\bk, \bk',\omega) & = & 
(2\pi)^2\delta^{(2)}(\bk-\bk')\,\delta_{pp'}\, r_{j;p}(\bk,\omega)\nonumber\\
 &+& R_{j;pp'}(\bk,\bk';\omega)\,H_{j}(\bk-\bk')
\label{reflectop}
\end{eqnarray}
The first line in this equation represents the zero-th order term with respect to corrugation,
that is also the specular reflection on a flat plate (with $r_{j;p}(\bk,\omega)$ the ordinary
specular reflection amplitude~\cite{lossy_cavities}).
The second line describes the first-order correction proportional to the 
Fourier component $H_j(\bk-\bk')$ of the profiles $h_j(\br)$, this 
Fourier component being able to induce a modification of the field wavevector
from $\bk$ to $\bk'.$

We then compute the correction of the Casimir energy $\delta E_\PP$ induced by the 
corrugations. At the lowest order, we have to use the scattering approach \cite{MLR2}
at second order in the corrugations, keeping only the crossed terms of the form $H_1\,H_2$
which have the ability to induce lateral forces. 
This means that the sensitivity function obtained below depends on the crossed correlation
between the profiles of the two plates, in contrast to the function which was calculated
for describing roughness correction in \cite{MLR,MLR2}.
The latter were depending on terms quadratic in $H_1$ or $H_2,$ and their evaluation
required that second order non specular scattering be properly taken into account.
Here, first order non specular amplitudes evaluated on both plates are sufficient.

Assuming for simplicity that the two plates are made of the same metallic medium, 
so that explicit reference to the index $j$ may be omitted in  
the reflection amplitudes from now on (otherwise the result is given by a trivial
extension), we obtain the second-order correction
\begin{equation}
\delta E_\PP=\int \frac{\dd^{2}\bk}{(2\pi )^{2}}{\cal G}(\bk)H_{1}(\bk)H_{2}(-\bk)  
\label{Epp}
\end{equation}
The non linear response function ${\cal G}(\bk)$ is given by
\begin{eqnarray}
&&{\cal G}(\bk)=-\hbar \int_{0}^{\infty }\frac{\dd\xi }{2\pi }\int 
\frac{\dd^{2}\bk'}{(2\pi )^{2}}b_{\bk',\bk'-\bk}(\xi )  
\label{F(k)3} \\
&&b_{\bk',\bk}=\sum_{p',p} \frac{e^{-(\kappa'+\kappa)L}
R_{p'p}(\bk',\bk;\xi) R_{pp'}(\bk,\bk';\xi) }
{\left(1-r_{p'}(\bk^{\prime})^2 \,e^{-2\kappa'L} \right)
\left(1-r_{p} (\bk)^2\,e^{-2\kappa L}\right)}\nonumber
\end{eqnarray}
Note that the integral over real frequencies $\omega$ has been replaced by an integral 
over imaginary frequencies $\xi,$ while the round-trip propagation factor 
between the plates is now expressed as $\exp(-2\kappa L),$ with 
$\kappa = \sqrt{k^2+\xi^2/c^2}.$  
For isotropic media, symmetry requires the response function
${\cal G}(\bk)$ to depend only on the modulus $k=|\bk|.$

Experiments with corrugated plates~\cite{chen} were corresponding to the simple case 
where uniaxial sinusoidal corrugations are imprinted on the two plates along the same direction, say the $y$ direction, and with the same wavevector $k \equiv 2\pi /\lambda_\C$
\begin{eqnarray}
h_{1} =a_1\cos \left(kx\right) \quad,\quad h_{2} =a_2\cos \left( k(x+b) \right) &&
\label{profiles}
\end{eqnarray}
The energy correction thus depends on the lateral mismatch $b$ between 
the corrugations of the two plates, which is the cause for the lateral 
force to arise.
Replacing the ill-defined $\left( 2\pi \right) ^{2}\delta ^{(2)}(0)$ by the area  
$A$ of the plates, we derive from (\ref{Epp})
\begin{equation}
\delta E_{\PP}=A\frac{a_1 a_2}{2} \cos (kb) {\cal G}(k)  \label{Epp1/2}
\end{equation}

The result of the PFA can be recovered from Eq.~(\ref{Epp1/2}) as the limiting case $k\rightarrow0$.  
PFA indeed correspond to long corrugation wavelengths, that is also nearly plane surfaces
so that the Casimir energy is obtained from the plane-plane energy  $E_\PP (z)$
by averaging the `local' distance $z=L-h_{1}-h_{2}$ over the surface of the plates~\cite{EPL}.
Expanding at second order in the corrugations, we thus obtain the following dependence
of the energy (as before we disregard terms in $a_1^2$ and $a_2^2$
because they do not depend on the relative lateral position)
\begin{equation}
\delta E_\PP=\frac{a_1a_2}{2} \cos (kb) \frac{\dd^2 E_\PP}{\dd z^2}
\quad,\quad[{\rm PFA}]
\label{PFA2}
\end{equation}
This reproduces (\ref{Epp1/2}) for large values of $\lambda_\C$ because the response function
${\cal G}(k)$ satisfies the condition 
\begin{equation}
\lim_{k\rightarrow 0} A{\cal G}(k)=\frac{\dd^2 E_\PP}{\dd z^2}
\label{PFA3}
\end{equation}
This property is ensured, for any specific model of the material medium, by the
fact that ${\cal G}$ is given for $k\rightarrow0$ by the specular limit of non 
specular reflection amplitudes (more details in~\cite{MLR2}).
For arbitrary values of $k$, the deviation from PFA is described by the ratio
\begin{equation}
\rho(k)=\frac{{\cal G}(k)}{{\cal G}(0)} \quad,\quad
\lim_{k\rightarrow 0} \rho(k)=1
\label{PFA4}
\end{equation}

\begin{figure}[ptb]
\centerline{\psfig{figure=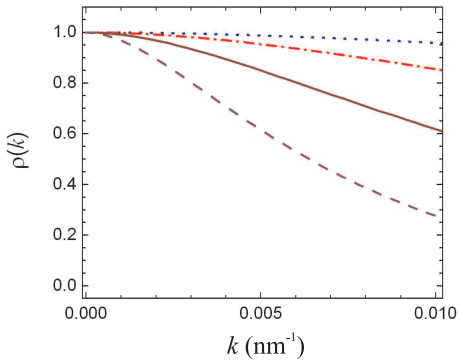,width=7cm}}
\caption{
Variation of $\rho$ versus $k$ with $\lambda_\P=136$nm and for  
$L=$ 50nm (dotted line), 100nm (dash-dotted line), 200nm
(solid line) and 400nm (dashed line).}
\label{ratio}
\end{figure}

In the following, we discuss explicit expressions of this ratio $\rho$ 
obtained from the general result~(\ref{F(k)3}) in the particular case of metallic mirrors
described by the plasma model. 
The dielectric function is thus given by $\epsilon =1+\omega _\P^2/\xi^2,$ with
the plasma wavelength and frequency related by $\lambda _\P=2\pi c/\omega_\P.$
For the numerical examples, we will take $\lambda_\P=136$nm,
corresponding to gold covered plates. 
The non-specular coefficients appearing in Eq.~(\ref{F(k)3}) are then evaluated from 
Ref.~\cite{MLR2}. 
The resulting function $\rho$ is plotted on Fig.~\ref{ratio} as a function of $k,$
for different values of the separation distance $L.$ 

\begin{figure}[ptb]
\centerline{\psfig{figure=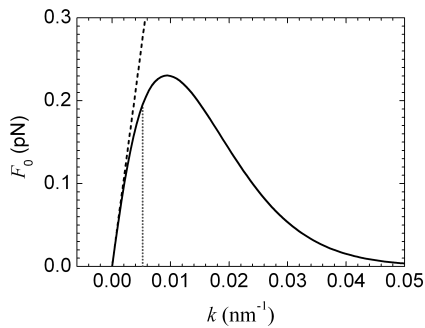,width=7cm}}
\caption{Lateral force amplitude for the plane-sphere setup, as a function of $k,$ 
with figures taken from~\cite{chen}. 
The experimental value $k=0.0052 {\rm nm}^{-1}$ is indicated by the vertical dashed line. }
\label{vsk}
\end{figure}

As expected, the PFA ($\rho=1$) is a better approximation at short distances. 
For $L=$50nm for example, the approximation is correct in the range 
$k\le 0.01 {\rm nm}^{-1}$ ({\it i.e.} $\lambda_\C\ge 628$nm)
covered by the plot in Fig.~\ref{ratio}.
For larger distances, PFA is found to {\it overestimate} the energy correction 
and, therefore, the lateral force.
For parameters $L=$200nm and $\lambda_\C = 1.2 \mu$m ($k=0.0052 {\rm nm}^{-1}$) 
close to the experimental figures of Ref.~\cite{chen},
we find the lateral force smaller by a factor $\rho = 0.84.$ 
In other words, the correction expected with respect to PFA calculations is of the order of $16\%$.

For still larger values of $k,$ the functions ${\cal G}(k)$ and $\rho(k)$ decay 
exponentially to zero. When the momentum transfer $k$ is much larger than $1/L$, the function 
$b_{\bk',\bk'-\bk}$ in Eq.~(\ref{F(k)3}) is proportional to the
exponentially small factor $\exp (-kL)$. 
If we also assume that $k\lambda_\P\gg 1,$ we find 
$
{\cal G}(k)=\alpha \,k\,\exp (-kL)
$
where the parameter $\alpha $ now depends on $\lambda_\P$ and $L$ only. 
This is in striking contrast with the behavior of the response function for stochastic 
roughness, which {\it grows} linearly with $k$ for large $k$ due to  
 the contribution of  the second-order reflection coefficients \cite{MLR2}. 
These coefficients do not contribute to the second-order lateral effect,
 which  is related to two first-order non-specular reflections at different plates,
separated by a one-way propagation with a modified momentum of the order of $k.$
The resulting propagation factor is, in the large-$k$ limit, 
 $\exp(-\kappa L)\approx\exp(-kL),$
thus explaining the exponential behavior. 

For intermediate values of $k,$ ${\cal G}(k)$ is also smaller than the roughness response function, although 
no simple analytical expression is available in the general case. Thus, the 
 exact calculation decreases the PFA value for the lateral effect, in contrast to the case of 
stochastic roughness correction where they increase the PFA value~\cite{MLR,MLR2}.

We now compare the theoretical expression of
the lateral Casimir force to realistic experiments and therefore
consider the plane-sphere (PS) geometry~\cite{chen} rather
than the plane-plane (PP) geometry as before. 
The PS result can be derived from the PP one by using PFA with validity conditions
more easily met than those which would be necessary for computing the corrugation effect. 
As a matter of fact, PFA may account for the curvature of the sphere as soon as 
the radius $R$ of the sphere is larger than the distance $R\gg L.$ Then any
interplay between curvature and corrugation is avoided provided that $RL\gg \lambda_\C^2.$ 
These two conditions are met in the experiment reported in~\cite{chen}, 
where $R=100 \mu$m, $\lambda_\C=1.2 \mu$m and $L\sim 200$nm. 
PFA is likely a very good approximation as far as curvature is concerned whereas, 
for the corrugation effect itself, a deviation from the PFA is expected 
since $L$ is not so much smaller than $\lambda_\C$.

Applying the PFA to describe the PS configuration, we obtain the energy correction  
$\delta E_\PS$ between the sphere and a plane at a distance of closest approach $L$ as an 
integral of the energy correction $\delta E_\PP$ in the PP geometry  
\begin{equation}
\delta E_\PS(L,b) = \int_{\infty}^{L} \frac{2\pi R\dd L'}{A} \delta E_\PP(L',b) 
\label{FPSsmooth}
\end{equation}
Then the lateral force is deduced by varying the energy correction
with respect to the lateral mismatch $b$ between the two corrugations.
This gives the lateral Casimir force in the PS geometry as
an average of 
\begin{eqnarray}
&&F^\lat_\PS(L,b)\equiv-\frac{\partial}{\partial b} E_\PS(L,b) =
\int_{\infty}^{L} \frac{2\pi R\dd L'}{A} F^\lat_\PP(L',b) \nonumber\\
&&F^\lat_\PP(L,b)\equiv-\frac{\partial }{\partial b} E_\PP(L,b)
\label{Forca lateral}
\end{eqnarray}
leading eventually with Eq.~(\ref{Epp1/2}) to
\begin{equation}
F^\lat_\PS = \pi a_1 a_2 \,k R \sin(kb) \int_{\infty}^{L} \dd L' {\cal G}(k,L')
\end{equation}
The force attains a maximal amplitude for $\sin(kb)=\pm1$, which is
easily evaluated in the PFA regime $k\rightarrow0$
where ${\cal G}(k)$ does not depend on $k$, so that $F^\lat_\PS$ scales as $k.$ 
However, as $k$ increases, the amplitude increases at a slower rate and then starts to 
decrease due to the exponential decay of ${\cal G}(k).$ 
Thus, for a given value of the separation $L,$ the lateral force reaches an optimum 
for a corrugation wavelength such that $kL$ is of order of unity, 
which generalizes the result obtained for perfect reflectors in~\cite{emig}. 
In Fig.~\ref{vsk}, we plot the force $F^\lat_\PS$ (for $\sin(kb)=1$) as a function of $k,$
with figures taken from the experiment of Ref.~\cite{chen}. 
We also use the values $a_1=59$nm and $a_2=8$nm of the amplitudes for measuring the force
as in~\cite{chen}, reminding however that our calculations are valid in the perturbative limit
$a_1,a_2\rightarrow0$. 

\begin{figure}[ptb]
\centerline{\psfig{figure=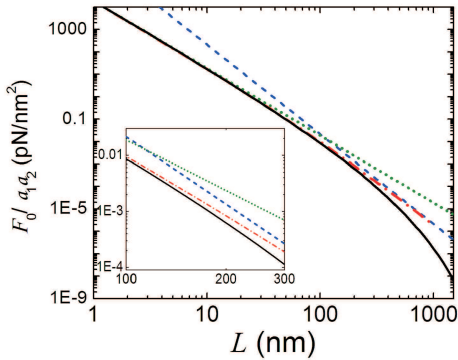,width=7cm}}
\caption{Plane-sphere geometry: $F^\lat_\PS/(a_1 a_2 \sin(kb))$ as a function of $L,$ 
with $k=0.0052 {\rm nm}^{-1}.$ The solid line is the exact result obtained in this paper.
Other lines represent various approximations: PFA (dotted-dashed)
and PFA combined with the 
 perfectly-reflecting (dashed) and plasmon (dotted) limits.
}
\label{vsL}
\end{figure}

The plot clearly shows the linear growth for small $k$ as well as the exponential decay
for large $k.$ 
The maximum force is at $k=0.009{\rm nm}^{-1}$ so that $kL\simeq2$.
The experimental value $k=0.0052 {\rm nm}^{-1}$ is indicated by the dashed line in 
Fig.~\ref{vsk}, and the force obtained as 0.20pN, well below the PFA result,
indicated by the straight line and corresponding to a force of 0.28pN.
This last value is very close to the number calculated by Ref. \cite{chen}, 
if we discount the correction due to higher order terms beyond the second order 
approximation considered in the present paper. 
Precisely, Ref. \cite{chen} finds a force of 0.32pN at $L=221$nm, with a relative correction 
due to higher powers of $1.21.$ 
Discounting this factor, the second-order force should be 0.26pN,
which overestimates the exact result by a factor of the order of 30\%.

In order to discuss the dependence of the force versus the distance $L$, we plot
$F^\lat_\PS/(a_1 a_2 \sin(kb))$ with log-log scales on Fig.~\ref{vsL}, 
with $\lambda_\C=1.2 \mu$m (solid line).
For very small values of $L,$ we find the $L^{-3}$ power law expected for the plasmon limit 
$L\ll\lambda_\P$ \cite{plasmon} combined with PFA (dotted line). 
We also show the PFA result for arbitrary $\lambda_P/L$ (dotted-dashed line) and 
the $1/L^4$ dependence resulting from the combination of PFA and perfectly-reflecting 
assumption (dashed line). Fitting in the range between 150 and 300 nm, we find the power law $L^{-4.1}$, in agreement with the experimental result~\cite{chen}.  
But Fig.~\ref{vsL} shows that the force decays faster for larger values of $L,$  
leading ultimately to an exponential decay.

These results indicate the way to be followed to perform an accurate comparison between
theory and experiment in a configuration where geometry plays a non trivial role, {\it i.e.}
beyond the PFA. Such a comparison should be pursued 
by looking at the lateral force between corrugated metallic plates.
This problem can neither be addressed by calculations performed with the help of 
PFA~\cite{chen,blagov}, nor with the assumption of perfect reflection~\cite{emig}. 
But it can be tackled by the technique presented in the present letter
which is valid for arbitrary relative values of $L,$ $\lambda_\C$ and $\lambda_\P$, provided that the corrugation 
amplitude is smaller than the other length scales.

One of us (P. A. M. N.) thanks CNPq and Instituto do Mil\^enio de Informa\c c\~ao Qu\^antica for partial financial support.

\end{document}